\documentclass[preprintnumbers,prd,onecolumn,floatfix,superscriptaddress, nofootinbib]{revtex4-2}

\usepackage{setspace}
\usepackage{graphicx}
\usepackage{subcaption}
\usepackage{epsfig}
\usepackage{bm}
\usepackage{amssymb}
\usepackage{float}
\usepackage{amsmath}
\usepackage{dcolumn}
\usepackage{cancel}
\usepackage[colorlinks]{hyperref}
\usepackage[usenames,dvipsnames]{color}
\usepackage{enumitem}

\hypersetup{ breaklinks=true, pdfstartview={FitH}, colorlinks=true, linkcolor=blue, citecolor=red, filecolor=magenta, urlcolor=blue, anchorcolor=green, linktocpage=true }

\def\Om{\Omega}

\def\doi{http://doi.org}

\newcommand{\be}{\begin{equation}}
\newcommand{\ee}{\end{equation}}
\newcommand{\beano}{\begin{eqnarray*}}
\newcommand{\eeano}{\end{eqnarray*}}
\newcommand{\ba}{\begin{eqnarray}}
\newcommand{\ea}{\end{eqnarray}}

\begin{document}

\title{ New parametrization of the dark-energy equation of state with a single parameter}
\author{J. K. Singh}
\email{jksingh@nsut.ac.in}
\affiliation{Department of Mathematics, Netaji Subhas University of Technology, New Delhi-110078, India}
\author{Preeti Singh}
\email{diasprity@gmail.com}
\affiliation{Department of Mathematics, Netaji Subhas University of Technology, New Delhi-110078, India} 

\author{Emmanuel N. Saridakis}
\email{msaridak@noa.gr}
\affiliation{National Observatory of Athens, Lofos Nymfon, 11852 Athens, Greece}
\affiliation{CAS Key Laboratory for Research in Galaxies and Cosmology,  
 University of Science and Technology of China, Hefei, Anhui 230026, China}
\affiliation{Departamento de Matem\'{a}ticas, Universidad Cat\'{o}lica del Norte, Avda.
Angamos 0610, Casilla 1280 Antofagasta, Chile}
   
\author{Shynaray Myrzakul}
\email{srmyrzakul@gmail.com}
\affiliation{ Ratbay Myrzakulov Eurasian International Centre for Theoretical Physics, Nur-Sultan 010009, Kazakhstan}

\author{Harshna Balhara}
\email{harshna.ma19@nsut.ac.in}
\affiliation{Department of Mathematics, Netaji Subhas University of Technology, New Delhi-110 078, India}

\begin{abstract}
We propose a novel dark-energy equation-of-state parametrization, with a single parameter $\eta$ that quantifies the deviation from $\Lambda$CDM cosmology. We first confront the scenario with various datasets, from  Hubble function  (OHD), Pantheon, baryon acoustic oscillations (BAO), and their joint observations, and we show that $\eta$ has a preference for a non-zero value, namely a deviation from $\Lambda$CDM cosmology is favored, although the zero value is marginally inside the 1$\sigma$ confidence level. However, we find that the present Hubble function value acquires a higher value, namely  $ H_0= 66.624^{+0.011}_{-0.013}~Km~ s^{-1} Mpc^{-1} $, which implies that the $H_0$ tension can be partially alleviated. Additionally, we perform a cosmographic analysis, showing that the universe transits from deceleration to acceleration in the recent cosmological past, nevertheless, in the future, it will not result in a de Sitter phase, since it exhibits a second transition from acceleration to deceleration.  Finally, we perform the Statefinder analysis. The scenario behaves similarly to the $ \Lambda$CDM paradigm at high redshifts, while the deviation becomes significant at late and recent times and especially in the future.

\end{abstract}
 
\pacs{98.80.-k, 95.36.+x, 98.80.Es}
   
\maketitle

\section{Introduction} 

The understanding of fundamental physics is challenged by the universe's rapid expansion, which is the most exciting subject in modern cosmology. This phenomenon indicates that, contrary to popular belief, the cosmos is expanding faster with time. It was initially discovered in the late $20^{th}$ century through observations of distant supernovae. Under the general relativistic framework, the first general class of explanation involves introducing new and exotic sectors in the universe content, under the umbrella term dark energy \cite{Copeland:2006wr, Cai:2009zp}. Extending the fundamental principles of gravity constitutes the second general class in which case the cause of acceleration is of gravitational origin \cite{DeFelice:2010aj, Capozziello:2011et, Cai:2015emx, Nojiri:2017ncd, CANTATA:2021ktz}.
 
Nevertheless, at the phenomenological level, both approaches can be quantified through the (effective) dark-energy  equation-of-state parameter  $w_x$. Thus, introducing various parametrizations of $w_x$ allows us to describe the universe's evolution and confront with observational datasets, to reveal the required dark-energy features to obtain agreement. Several researchers have embraced a phenomenological strategy for modeling the equation of state (EoS) parameter, representing it as a function of redshift, denoted as $\omega=\omega(z)$ \cite{Castillo-Santos:2022yoi}, or equivalently as a function of the cosmic scale factor, $\omega=\omega(a)$ \cite{Liu:2008vy, Perkovic:2020mph}. This methodology streamlines the examination of dark energy (DE) dynamics and enables the exploration of physically significant functions. However, during data fitting, it becomes necessary to simplify the parameterization of the equation of state, $\omega(z)$, and subsequently restrict the evolution of $\omega(z)$ based on the parameters we've introduced in our parametrization.
In particular, starting from the simple cosmological constant, a large number of parametrizations have been introduced  in the literature, involving one 
parameter \cite{Gong:2005de,Yang:2018qmz},  
  two parameters, such as the
Chevallier-Polarski-Linder (CPL) parametrization 
\cite{Chevallier:2000qy,Linder:2002et}, 
the Linear parametrization 
\cite{Cooray:1999da,Astier:2000as,Weller:2001gf}, the
Logarithmic parametrization \cite{Efstathiou:1999tm}, the 
Jassal-Bagla-Padmanabhan 
parametrization (JBP) \cite{Jassal:2005qc},  the
Barboza-Alcaniz (BA) parametrization \cite{Barboza:2008rh},  $ H_0 $ problem at low reshift \cite{Banerjee:2020xcn, Lee:2022cyh, Ma:2011nc,Nesseris:2004wj, Linder:2005ne, Feng:2004ff, Zhao:2006qg, Nojiri:2006ww, Saridakis:2008fy,Dutta:2009yb,Lazkoz:2010gz, Feng:2011zzo,Saridakis:2009pj, DeFelice:2012vd,Saridakis:2009ej, Feng:2012gf,Basilakos:2013vya, Pantazis:2016nky, DiValentino:2016hlg, Chavez:2016epc,Zhao:2017cud, Yang:2017amu, DiValentino:2017zyq, DiValentino:2017gzb, Yang:2017alx, Pan:2017zoh,Pan:2019jqh,Pan:2019gop,Singh:2022jue, 
Singh:2022nfm}, etc. Additionally, note that one can impose the parametrization at the deceleration 
parameter level \cite{Cunha:2008ja, Akarsu:2013lya, Xu:2009zza}, at equation-of-sate (EoS) parameter level \cite{Colgain:2021pmf} or even at the Hubble parameter level \cite{Singh:2018xjv, Singh:2019fpr, Nojiri:2015wsa, Nagpal:2019vre}.
 
In the present manuscript, we propose a novel dark-energy equation-of-state 
parametrization, with a single parameter $\eta$ that quantifies the deviation 
from $\Lambda$CDM cosmology. Additionally, under this scenario, dark energy 
behaves like a cosmological constant at high redshifts, while the deviation 
becomes significant at low and recent redshifts, especially in the future. 
Finally, for  $ \eta=0 $ we recover $ \Lambda$CDM cosmology completely. As we will 
see, apart from being capable of fitting the data, the new parametrization can 
partially alleviate the $ H_0 $ tension too, since it leads to a $ H_0 $ value in 
between the Planck one and the one from direct measurements \cite{Abdalla:2022yfr, DiValentino:2022uvj}. Recently, several authors have done so many remarkable works regarding the measurements of $ H_0 $ obtained from cosmological probes for different redshifts such as a Bias-free Cosmological Analysis with Quasars alleviating $ H_0 $ tension \cite{Lenart:2022nip}; a new statistical insights and cosmological constraints consists of Gamma-Ray Bursts, Quasars, Baryonic Acoustic Oscillations, and Supernovae Ia \cite{Dainotti:2023ebr}; reduced uncertainties up to $ 43\% $ on the Hubble constant and the matter density with the SNIa with a new statistical analysis \cite{Dainotti:2021pqg}; on the Hubble constant tension in the SNIa Pantheon sample \cite{Schiavone:2022shz}; on the evolution of the Hubble constant with the SNIa Pantheon Sample and Baryon Acoustic Oscillations: a feasibility study for GRB-cosmology in $ 2030 $ \cite{Dainotti:2022bzg}; $ f(R) $ gravity in the Jordan Frame as a paradigm for the Hubble tension, in which the authors provides a subsequent interpretation of the results through an effective Hubble constant that evolves with the redshift in a $ f(R) $ modified gravity theory in the Jordan frame \cite{Schiavone:2022wvq}. 
  
The article is organized as follows. In Section \ref{model} we present the novel dark-energy parametrization. Then in Section \ref{constraints} we perform a detailed confrontation with observations, namely with Hubble function (OHD), Pantheon, and baryon acoustic oscillations (BAO) data. In Section \ref{Cosmographicanalysis}  we perform a cosmographic analysis and we apply the Statefinder diagnostics. Finally,  Section \ref{Conclusions} is devoted to the conclusions. Lastly, the details of the various datasets and the corresponding fitting procedure is given in the appendix.
 
\section{New single-parameter equation-of-state parametrization}
\label{model}

In this section we first briefly review the basic equations of any cosmological 
scenario, and then we introduce the new parametrization for the dark-energy  
equation of state, with just a single parameter.
We consider the usual  homogeneous and isotropic  
Friedmann-Robertson-Walker (FRW)   metric
\begin{eqnarray}
{\rm d}s^2 = -{\rm d}t^2 + a^2 (t) \left[\frac{{\rm d}r^2}{1-Kr^2} + r^2 
\left(d 
\theta^2 
+ \sin^2 \theta d \phi^2\right)\right],
\end{eqnarray}
with $a(t)$    the scale factor  and $ K $ the spatial-curvature parameter, 
($K=0,-1,+1$ for spatially flat, open and closed universe, 
respectively). Furthermore, we consider that the universe is filled with 
baryonic and dark matter,  radiation  as well 
as the effective dark-energy fluid. Hence, the Friedmann equations that 
determine the background evolution of the Universe are
 \begin{eqnarray}
H^2 + \frac{K}{a^2} &=& \frac{8\pi G}{3} \rho_{tot},\label{efe1}\\
2\dot{H} + 3 H^2  + \frac{K}{a^2} &=& - 8 \pi G\, p_{tot}\label{efe2},
\end{eqnarray}
where $G$ is the gravitational constant and $H=\dot{a}/a$ the Hubble function, 
with dots marking time derivatives. The total energy density and pressure   
are thus given as 
$\rho_{tot} = \rho_r +\rho_b 
+\rho_c 
+\rho_x$ and $p_{tot} = p_r + 
p_b + p_c + p_x$, where  the subscripts $r,\; b,\; c,\; x$ stand respectively 
for
radiation, baryon, cold dark matter, and dark energy. As usual, and without 
loss of generality, we  focus on the spatially flat 
case and therefore in the following we impose  $K=0$.
Finally, assuming that the various sectors do not interact 
mutually we deduce that they are separately conserved, following the 
conservation equations
\begin{eqnarray}\label{cons}
\dot{\rho}_i + 3 H (1 +w_i ) \rho_i = 0,
\end{eqnarray} 
where $i \in \{ r, b, c, x\}$. In the above expression, we have introduced the 
equation-of-state parameter of each fluid as $p_i \equiv w_i \rho_i$ which yields
\begin{equation}\label{eos}
w_i= -\left( 1+ \frac{a }{3\rho_{i}}  \frac{ d \rho_i}{da}\right) .
\end{equation}

We proceed by providing for completeness the evolution equations of the 
Universe at the perturbation level. In synchronous gauge,  the perturbed metric 
reads as
\begin{eqnarray}
\label{perturbed-metric}
ds^2 = a^2(\tau) \left [-d\tau^2 + (\delta_{ij}+h_{ij}) dx^idx^j  \right],
\end{eqnarray}
with $\tau $ the conformal time, and  where $\delta_{ij}$  and $h_{ij}$ denote the unperturbed and the perturbed metric parts (with $h = h^{j}_{j}$ the trace). Perturbing additionally the universe fluids and transforming to the Fourier space we finally extract \cite{Mukhanov, Ma:1995ey, Malik:2008im}:
  \begin{eqnarray}
&&\delta'_{i}  = - (1+ w_{i})\, \left(\theta_{i}+ \frac{h'}{2}\right) - 
3\mathcal{H}\left(\frac{\delta p_i}{\delta \rho_i} - w_{i} \right)\delta_i - 9 
\mathcal{H}^2\left(\frac{\delta p_i}{\delta \rho_i} - c^2_{a,i} \right) (1+w_i) 
\frac{\theta_i}
{{k}^2}, \label{per1} \\
&&\theta'_{i}  = - \mathcal{H} \left(1- 3 \frac{\delta p_i}{\delta 
\rho_i}\right)\theta_{i} 
+ \frac{\delta p_i/\delta \rho_i}{1+w_{i}}\, {k}^2\, \delta_{i} 
-{k}^2\sigma_i,\label{per2}
\end{eqnarray}
with primes denoting conformal-time  derivative and with 
$\mathcal{H}= 
a^{\prime}/a$   the conformal Hubble function, and where ${k}$ is the mode 
wave number. Moreover,  $\delta_i = 
\delta \rho_i/\rho_i$ stands for the over density   of the $i$-th fluid,    
$\theta_{i}\equiv i k^{j} v_{j}$ marks the divergence of the $i$-th fluid 
velocity, and  
$\sigma_i$ is the corresponding anisotropic stress.
Lastly,  
$c_{a,i}^2 = \dot{p}_i/\dot{\rho}_i$ is the adiabatic sound speed   given as $ 
c^2_{a,i} =  w_i - 
\frac{w_i^{\prime}}{3\mathcal{H}(1+w_i)}$.
 
Let us now introduce the new dark-energy parametrization. As usual, knowing the 
equation of state of a fluid allows us to extract its time-evolution by solving 
the equation (\ref{cons}). For radiation, we have $ w_r=1/3 $ and 
thus we obtain  $\rho_r =\rho_{r_0}\, a^{-4}$ (setting the scale factor at 
present to 1), while for the baryonic and dark matter, we have $ w_b=w_c=0 $, which 
leads to $\rho_b = \rho_{b_0} \, a^{-3}$ and $\rho_c = \rho_{c_0}\, a^{-3}$, 
where  $\rho_{i_0}$  stands for the present density value of the  $i$-th fluid. 
Concerning the equation-of-state parameter of the dark-energy sector, since 
it is unknown, as we mentioned in the Introduction one can consider various 
parametrizations. Focusing on the barotropic fluid sub-class we consider that 
it is a function of time only, or equivalently of the scale factor $ a $, i.e. 
$ w_{x}(a) $. Hence, the solution of the EoS equation (\ref{eos}) leads 
to 
\begin{eqnarray}\label{de-evol}
\rho_{x}(a)=\rho_{x_0} \, a^{-3}\exp\left[
-3\int_{1}^{a}\frac{w_{x}\left(  a'\right)  }{a'}\,da'
\right].
\end{eqnarray}

In this work, we consider a novel dark-energy equation-of-state parameter (\ref{eos}) follows: 
\begin{equation}\label{24}
w_{x}(a)=-1+\frac{a^{-\eta}\,e^{-2\eta a}\,\eta\, \arctan a^{-\eta}}{3(1+a^{-2\eta})},
\end{equation} 
 where $\eta$ is the single parameter.  
Hence, introducing for convenience the redshift $ z $ as the independent variable (where $a^{-1}=1+z$) the above relation becomes
\begin{equation}\label{newEos}
w_{x}(z)=-1+\frac{(1+z)^{\eta}\,e^{ \frac{-2\eta}{(1+z)}}\,\eta\, \arctan (1+z)^{\eta}}{3(1+(1+z)^{2\eta})}.
\end{equation} 

Relation (\ref{newEos}) is the parametrization that we propose, and in the case $\eta=0$ we recover $\Lambda$CDM concordance model, where $w_{x}=-1$ and $\rho_{x}=\rho_{x_0}=const$ but in the general case, the parameter $\eta$ quantifies the deviation from $\Lambda$CDM scenario. However, note that for general $\eta$, for large redshifts, i.e. for $z\rightarrow\infty $, we acquire $w_{x} \rightarrow-1$, which implies that the deviation from $\Lambda$CDM scenario disappears in this regime, and thus Big Bang Nucleosynthesis bounds are immediately satisfied.
 
Inserting the above parametrization in the  first Friedmann equation 
(\ref{efe1}) we obtain
 \begin{equation}\label{20}
H= H_0 \sqrt{\left[ 
(\Omega_{b_0}+\Omega_{c_0})(1+z)^3+\Omega_{r_0}(1+z)^4+ \Omega_{
x_0} 
e^{\frac{\eta z}{1+z}} \frac{\arctan (1+z)^{\eta}}{\arctan 1}\right] },
\end{equation}
with $H_0$ the present value of the Hubble parameter, and where we have 
introduced the present values of the density parameters $\Omega_{i_0}\equiv  
\frac{ 8\pi G}{ 3H^2} \rho_{i_0}$ (hence the present value of the total matter 
density parameter is $ {\Omega_{m_0}}\equiv {\Omega_{b_0}}+{\Omega_{c_0}}$). 
This expression allows us to investigate 
the cosmological evolution in detail, and confront it with observational 
datasets. Actually, expression (\ref{20}), which is a simple deviation from $\Lambda$CDM cosmology only at small redshifts, while it recovers $\Lambda$CDM scenario at high redshifts, was the motivation behind parametrization (\ref{24}).
 
Lastly, from parametrization (\ref{newEos}) and the corresponding Hubble function (\ref{20}), we can straightforwardly calculate various quantities. In 
particular, the 
 deceleration parameter $ 
q=-1-\dot{H}{H^{-2}}$
 is given by
\begin{equation}\label{23}
q(z) = -1+\frac{\Big[3(1+z)^4\Omega_{m_0}+\frac{4\eta\Omega_{x_0}(1+z)^2 
e^\frac{\eta z}{1+z}\arctan (1+z)^{\eta-1}}{\pi(z^2+2z+2)}+\frac{4 \eta 
\Omega_{x_0} e^\frac{\eta 
z}{1+z}\arctan (1+z)^{\eta}}{\pi}\Big]}{2(1+z)\Big[(1+z)^3 
\Omega_{m_0}+\frac{4 \Omega_{x_0}e^\frac{\eta 
z}{1+z}\arctan (1+z)^{\eta}}{\pi}\Big]},
\end{equation}
while the higher-order cosmographic parameters 
    \cite{bolo}  read as
\begin{eqnarray} \label{39}
&&j=-q+2q(1+q)+(1+z)\frac{dq}{dz},\\
&&s=j-3j(1+q)-(1+z)\frac{dj}{dz},\\
&&l=s-4s(1+q)-(1+z)\frac{ds}{dz},\\
&&m=l-5l(1+q)-(1+z)\frac{dl}{dz}.
\label{39f}
\end{eqnarray}
Similarly, for the matter and dark-energy density parameters   we obtain
\begin{equation}\label{26}
\Omega_{m}(z)= \frac{1}{1+\frac{4\Omega_{x_0}e^{\frac{\eta 
z}{1+z}}\arctan (1+z)^\eta}{\pi\Omega_{b_0}(1+z)^3}},
\end{equation}
and
\begin{equation}\label{27}
\Omega_x(z)=\frac{1}{1+\frac{ \pi\,\Omega_{b_0} e^{\frac{-\eta 
z}{1+z}}[(1+z)^3 \arctan (1+z)^{-\eta}]}{4\,\Omega_{x_0}}}.
\end{equation}
 
\section{Observational constraints}
\label{constraints}

In the previous section, we proposed a new parametrization for the 
dark-energy equation of state, given by  (\ref{newEos}), which has a single 
parameter, namely $\eta$. In this section, we perform a detailed confrontation 
with various datasets \cite{ Shaily:2024nmy, Singh:2024aml, Singh:2024kez, Balhara:2023mgj, Riess:1998dv}, focusing on the bounds of $\eta$. In particular, we will use data from: (i) Hubble function observations (OHD) with 77 data points 
\cite{Shaily:2022enj}, (ii)  Pantheon with 1048 data points \cite{Pan-STARRS1:2017jku}, and   
(iii) baryon acoustic oscillations (BAO). The details of the datasets and the 
corresponding methodology are given in the Appendix. In our analysis we use the following priors: $H_0 \in [66,70]$, $\Omega_{m_0} \in [0.1,0.4]$, $\Omega_{x_0} \in [0.6, 0.8] $ and $\eta \in [0,1]$.

\begin{figure}
[ht]
\centering
	 { \includegraphics[scale=0.7]{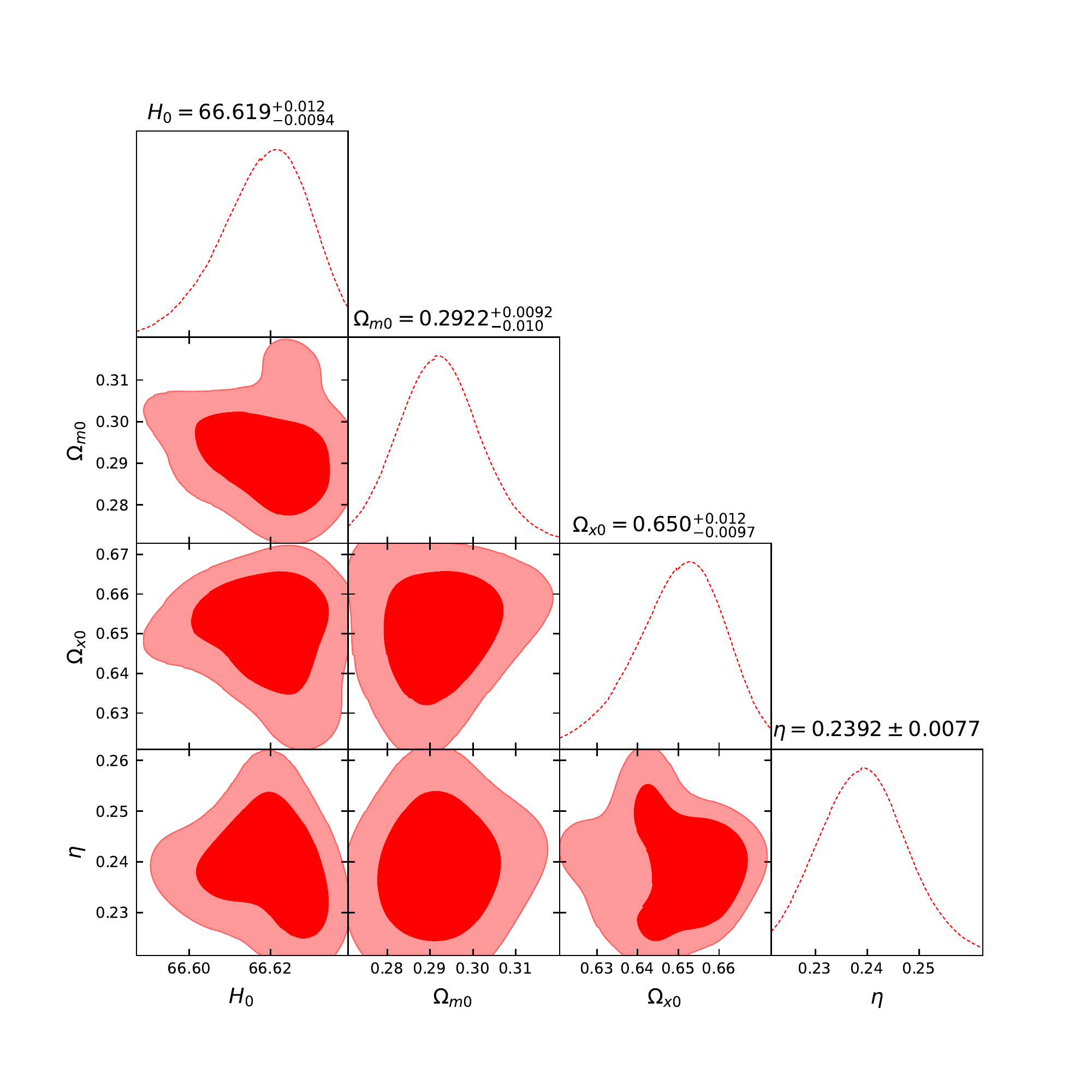}} 
		\caption{{\it{   The likelihood contours, with $ 1\sigma $ and $ 2\sigma $ confidence levels, for $ H(z) $ dataset.}}}
\label{Hz}
\end{figure}

\begin{figure}
[ht]
\centering
	 { \includegraphics[scale=0.7]{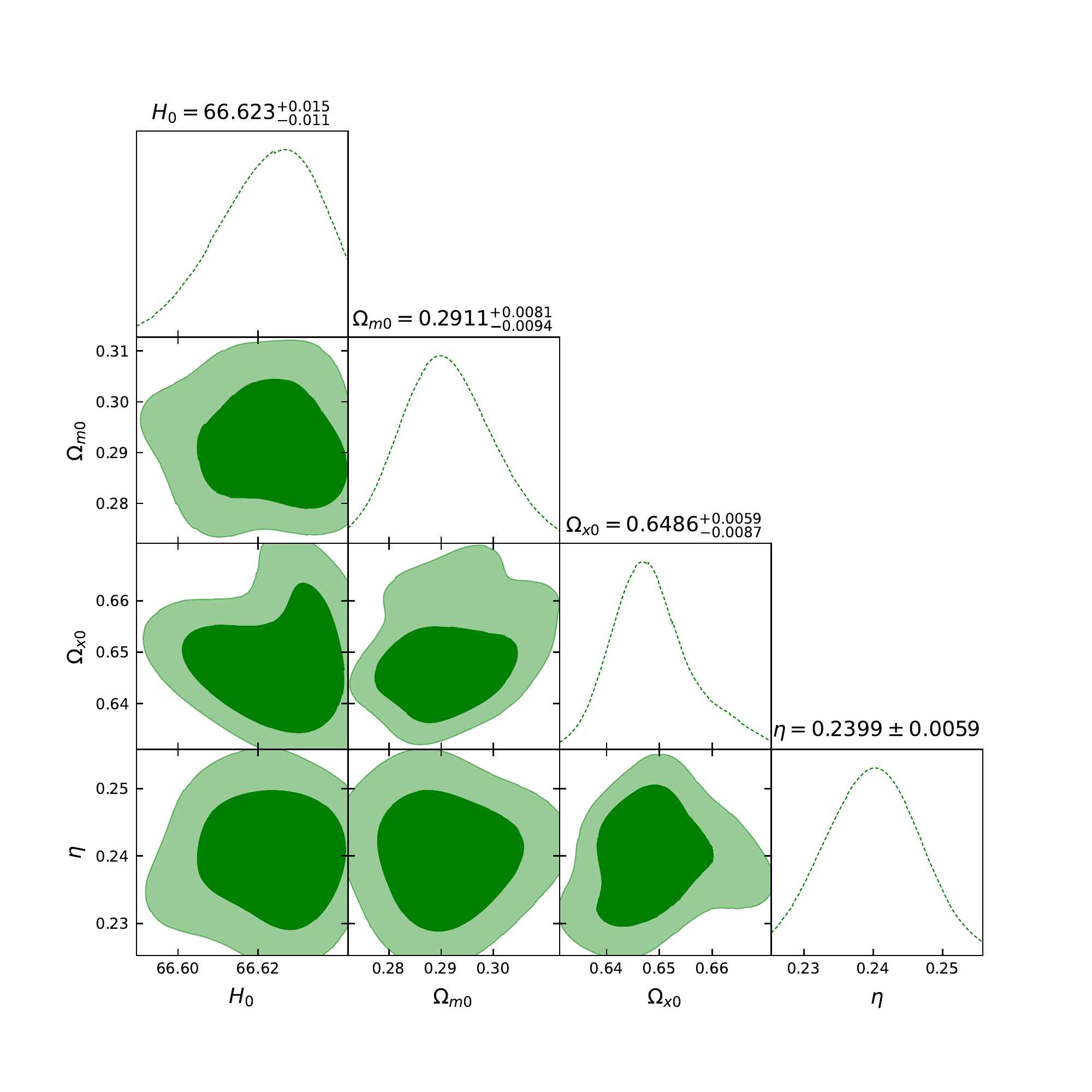}} 
		\caption{{\it{   The likelihood contours, with $ 1\sigma $ and $ 2\sigma $ confidence levels, for $ Pantheon $ dataset.}}}
\label{Pantheon}
\end{figure}

\begin{figure}
[ht]
\centering
	 { \includegraphics[scale=0.70]{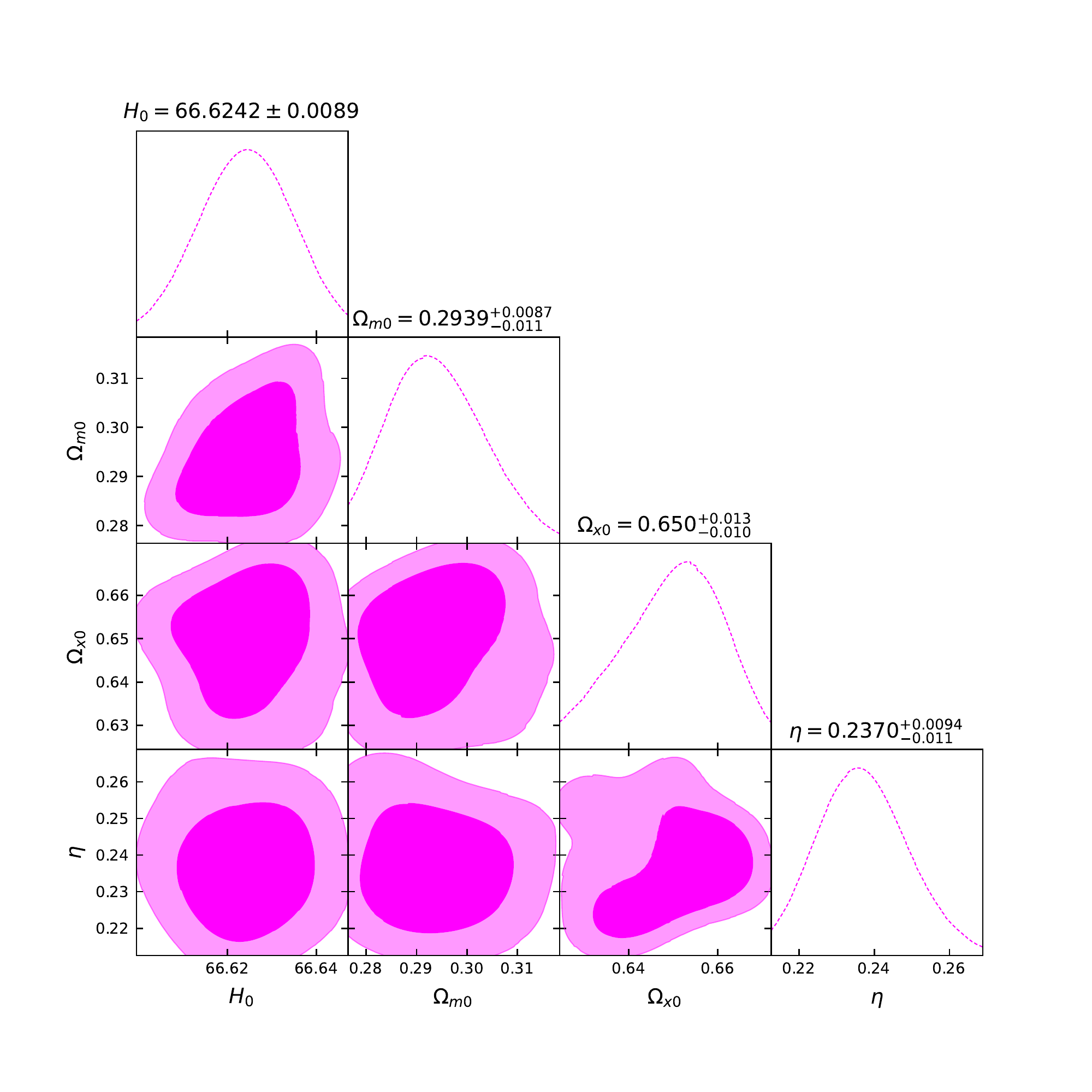}} 
		\caption{{\it{   The likelihood contours, with $ 1\sigma $ and $ 2\sigma $ confidence levels, for $ BAO $ data.}}}
\label{BAO}
\end{figure}

\begin{figure}
[ht]
\centering
	 { \includegraphics[scale=0.70]{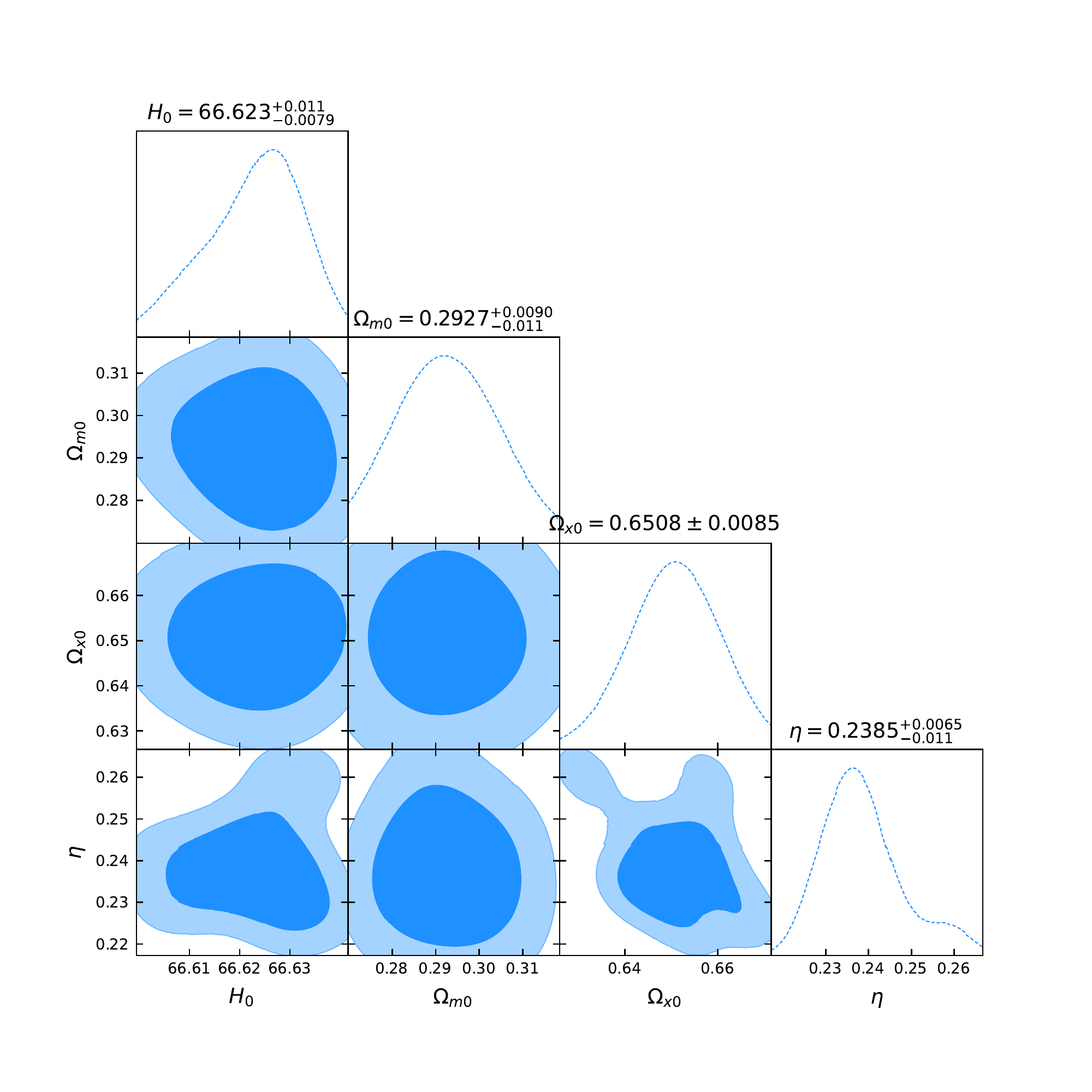}} 
		\caption{{\it{   The likelihood contours, with $ 1\sigma $ and $ 2\sigma $ confidence levels, for $ joint $ data.}}}
\label{Joint}
\end{figure}

Let us now present the constraints we obtain after applying the above formalism 
and datasets in the Friedmann equations at hand, focusing on the new model 
parameter $\eta$.   In Figs. \ref{Hz},\ref{Pantheon},\ref{BAO},\ref{Joint} we 
present the likelihood contours with $ 1\sigma $ and $ 2\sigma $ confidence 
levels, around the best-fit values. Additionally,  in Table \ref{tabparm1} we summarize the obtained results. Finally, in Table \ref{tabparm2}  we summarize other cosmological parameters, such as the density parameters and the deceleration 
parameter, the equation-of-state parameters, and the transition redshift.  

\begin{table}[ht]
\caption{ Summary of the observational constraints on the model parameters $H_0$, $ \Om_{m0}$, $\Om_{x0}$ and $\eta $ from various datasets.}\label{tabparm1}
\begin{center}
\begin{tabular}{l c c c c r} 
\hline\hline
{Dataset} &  $ H_0$ {\footnotesize(km/s/Mpc)} \,\,\,   &  \,\,\,   $ \Om_{m0} $ \,\,\, & $\Om_{x0}$ \,\,\, & $\eta$ \,\,\,
\\
\hline     
{\scriptsize {$ H(z) $ }} { \scriptsize{(77 points data)}}  &    $ 66.619^{+0.012}_{-0.0094} $    &   $ 
0.2922^{+0.0092}_{-0.010} $   &   $ 0.650^{+0.012}_{-0.0097} $ & $ 0.2392^{+0.0077}_{-0.0077}$
\\
{\scriptsize { $ \!\! Pantheon $  }}  &  $ 66.623 
^{+0.015}_{-0.011} $  &  $ 0.2911 
^{+0.0081}_{-0.0094}$  &  $ 0.6486 ^{+0.0059}_{-0.0087}$ & $0.2399^{0.0059}_{0.0059} $
\\
{\scriptsize { $ BAO $ }} &  $ 66.6242^{+0.0089}_{-0.0089} $  &  $ 
0.2939^{+0.0087}_{-0.011} $   &  $ 0.650 ^{+0.013}_{-0.010}$ & $0.2370^{+0.0094}_{-0.011}$
\\
{\scriptsize {$ H(z) $ + $ Pantheon$ + $ BAO $ } } &  $ 66.623^{+0.011}_{-0.0079} $  &  $ 
0.2927^{+0.0090}_{-0.011} $  &  $ 0.6508^{+0.0085}_{-0.0085} $ & $0.2385^{+0.0065}_{-0.011}$
\\
\hline\hline  
\end{tabular}    
\end{center}
\end{table}

\begin{table} 
\caption{ Summary of the constraints on the deceleration 
parameter $ q $, the transition redshift $ z_{tr} $ and the current value of the 
dark-energy equation-of-state parameter $ w_{x_0} $ 
. }\label{tabparm2}
\begin{center}
\begin{tabular}{l c c c c c r} 
\hline\hline
{Dataset} & $ q_0 $  &  $ z_{tr} $  &      $ w_{x_0} $   & \ \ \ \ \ \ \  \
\\
\hline
{ \scriptsize{ $ H(z)$}}
{ \scriptsize{(77 points data)}}    &   $ -0.449196 $  &  $ \ \ 
\
\simeq 0.7804\ \ 
\ $     & $ -0.704911 $   
\\
{\scriptsize{ $ \ Pantheon $ }}  &  $ -0.449152 $ &  $ \ 
\ 
\
\simeq 0.7811\ \ 
\ $ &   $ -0.704047 $ 
\\
{\scriptsize{$ \ \, BAO $ } }
 &  $ -0.449141 $  &  $\ \ 
\ \simeq 
0.7759\ \ 
\ 
$  &    $ -0.707625 $ 
\\
{ \scriptsize{$ H(z) $ + $ Pantheon$}} 
{ \scriptsize{ + $ BAO $ }}  &  $ -0.449532 $  &   $\ \ 
\ \simeq 
0.7804\ \ 
\ $  &    $ -0.705775 $ 
\\
\hline\hline  
\end{tabular}    
\end{center}
\end{table}

As we observe, the new model parameter  $ \eta$ that quantifies the deviation from  $\Lambda$CDM cosmology has a preference for a non-zero value, although zero is marginally inside the 1$\sigma$ confidence level. Concerning $ H_0 $ we observe that we obtain a higher value compared to $\Lambda$CDM scenario, although a bit lower than the direct measurements $ H_0= (73.04±1.04)~Km~ s^{-1} Mpc^{-1} $ at $68 \% $ CL, based on the Supernovae calibrated by Cepheids \cite{Abdalla:2022yfr}, which implies that the new dark-energy parametrization at hand can partially alleviate the $ H_0 $ tension. This is an additional result of the present work.

\textbf{Fig. \ref{Pantheon} has been plotted using the Pantheon data points comprising a full covariance matrix}. In the case of Fig. \ref{Pantheon}, we notice that the shape of the contour is not oval i.e., the posterior distribution for eta behaves as a bimodal distribution and therefore we need to test the convergence for the Pantheon dataset. The Gelman-Rubin convergence test \citep{Gelman:1992zz} is one of the statistical tools in Bayesian inference that is widely used to access the convergence of the chains. The test is based on the idea that multiple MCMC chains with different starting points should converge to the same posterior distribution if they have been run for long enough, that is, after several steps. Essentially, this test evaluates, for each parameter of the discussed model, the term called potential scale reduction $\hat R$, which is the ratio between the variance $\rm W$ within a chain and the variance $\rm Var(\theta)$ among the chains.
\begin{equation}
    \hat R=\sqrt{\frac{\rm Var(\theta)}{W}}.
\end{equation}
Also, in the likelihood contour of the Pantheon dataset, there is a possibility of $\pm 1.2$ for $ H_0 $ value and $\pm 0.29$ for $\alpha$ value within $100$ steps of burn-in which were used to run the MCMC chains.

The maximum Gelman–Rubin diagnostic across the model parameters is labeled as Max Gelman–Rubin $\hat R $ in the header and is less than $1.2$.  \citep{Gelman:1992zz} and \citep{Brooks:1998} suggest that diagnostic $\hat R $ values greater than $ 1.2$ for any of the model parameters should indicate non-convergence. The contour plots in the plane $ \eta \,-\,  H_{0} $ with $ 1\sigma $ and $ 2\sigma $ errors are given in Fig.  \ref{Pantheon}, and the corresponding best-fit values of $H_0$, and $\eta $ for different datasets are given in Table \ref{tabparm1}.

\section{Cosmographic analysis and Statefinder diagnostic}
\label{Cosmographicanalysis}

In this section, for completeness, we perform the cosmographic analysis of the 
cosmological scenario with the new dark-energy parametrization, and we apply 
the Statefinder diagnostic. For simplicity, we neglect the radiation sector.

\begin{figure}
[ht]
\centering
	 { \includegraphics[scale=0.35]{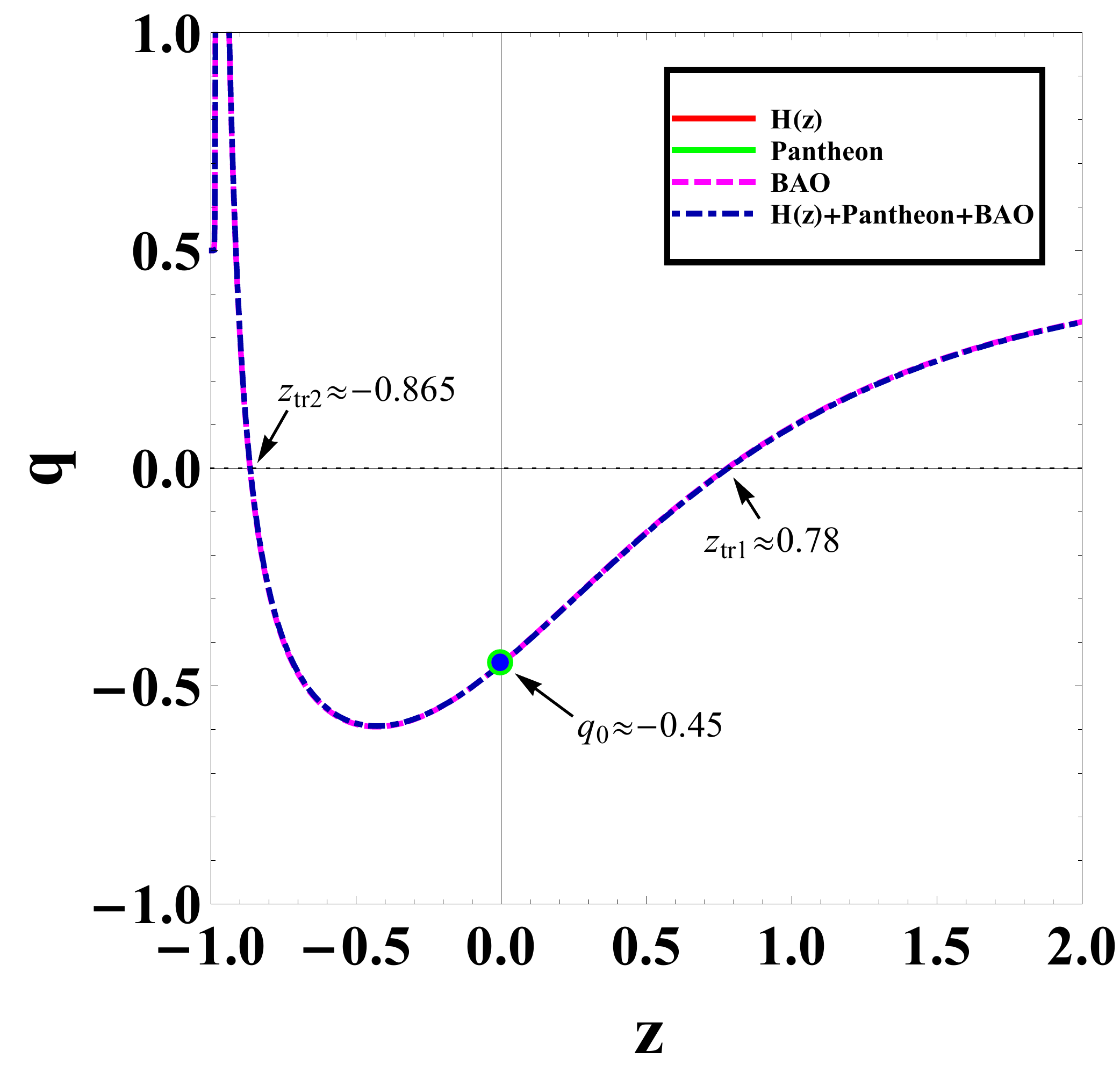}} 
		\caption{{\it{  The evolution of  the deceleration parameter $ q $ in 
terms of the redshift $z$, using the 
best-fit values of the model parameters given in Tables 
\ref{tabparm1}-\ref{tabparm2}. The red dots mark the present values.}}}
\label{qz}
\end{figure}

Let us start with the deceleration parameter given in  (\ref{23}). Using the best-fit values of the model parameters given in Tables \ref{tabparm1}-\ref{tabparm2}, we plot $q(z)$ in Fig. \ref{qz}. As we can see, we obtain the transition from deceleration to acceleration at the transition redshift $ z_{tr} $ in agreement with observations.  However, it is interesting to note that the novel parametrization at hand will lead to a second transition in the future, at redshift $ z_{tr_2}$, from acceleration to deceleration (at around $z_{tr_2}\approx-0.9$).  

\begin{figure} [ht]
 { \includegraphics[scale=0.28]{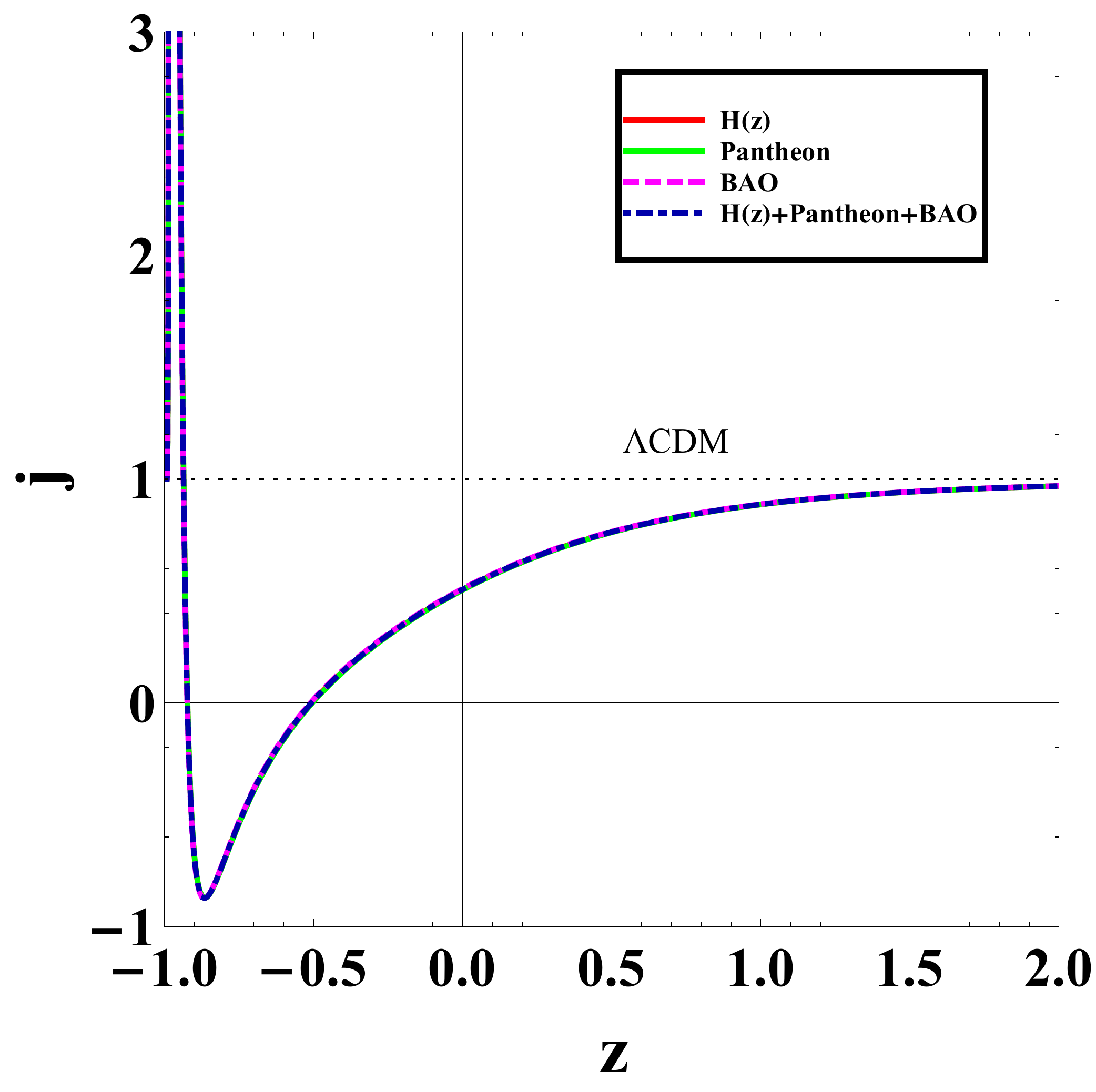}}
	 { \includegraphics[scale=0.28]{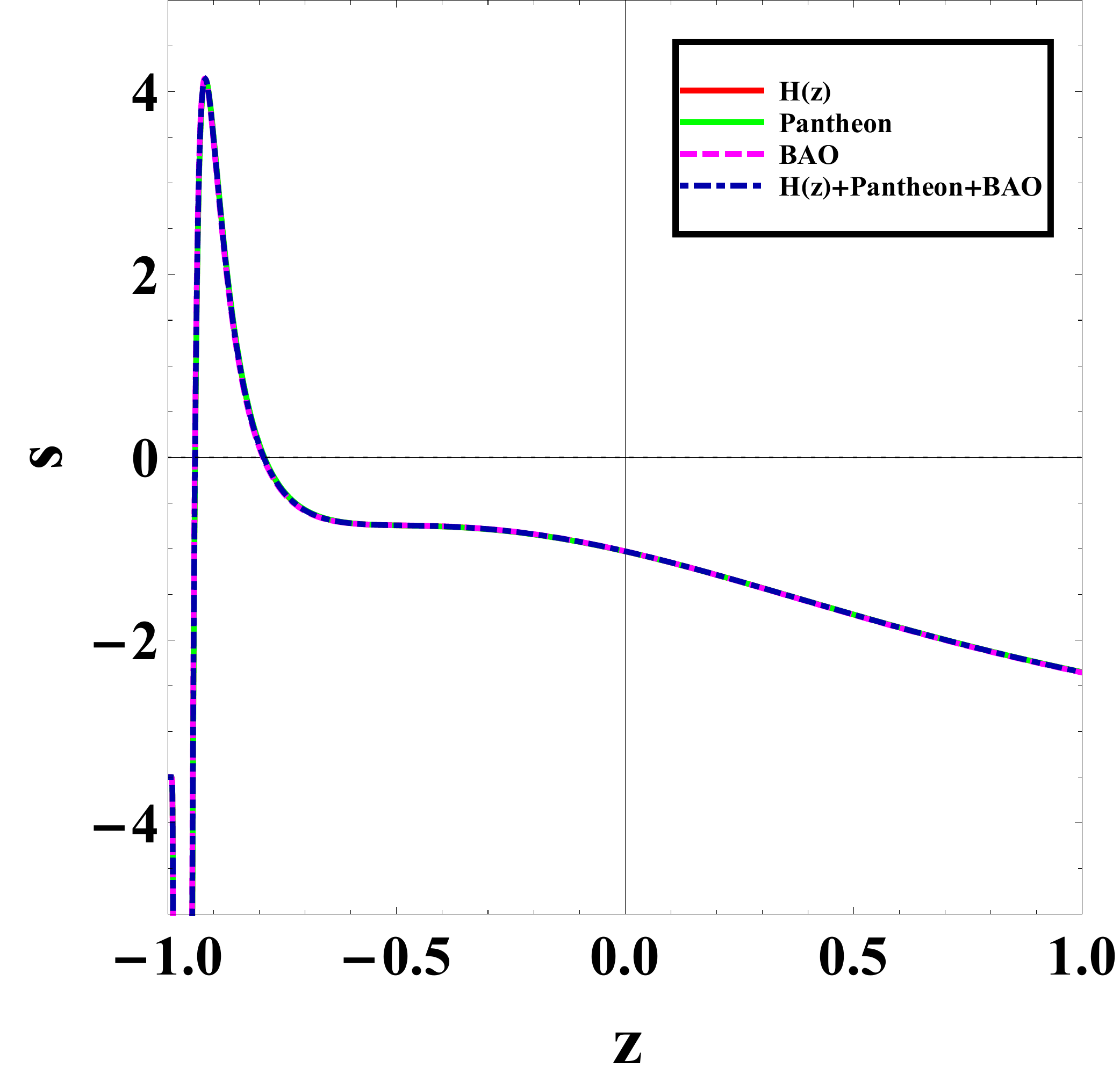}}\\
	 { \includegraphics[scale=0.28]{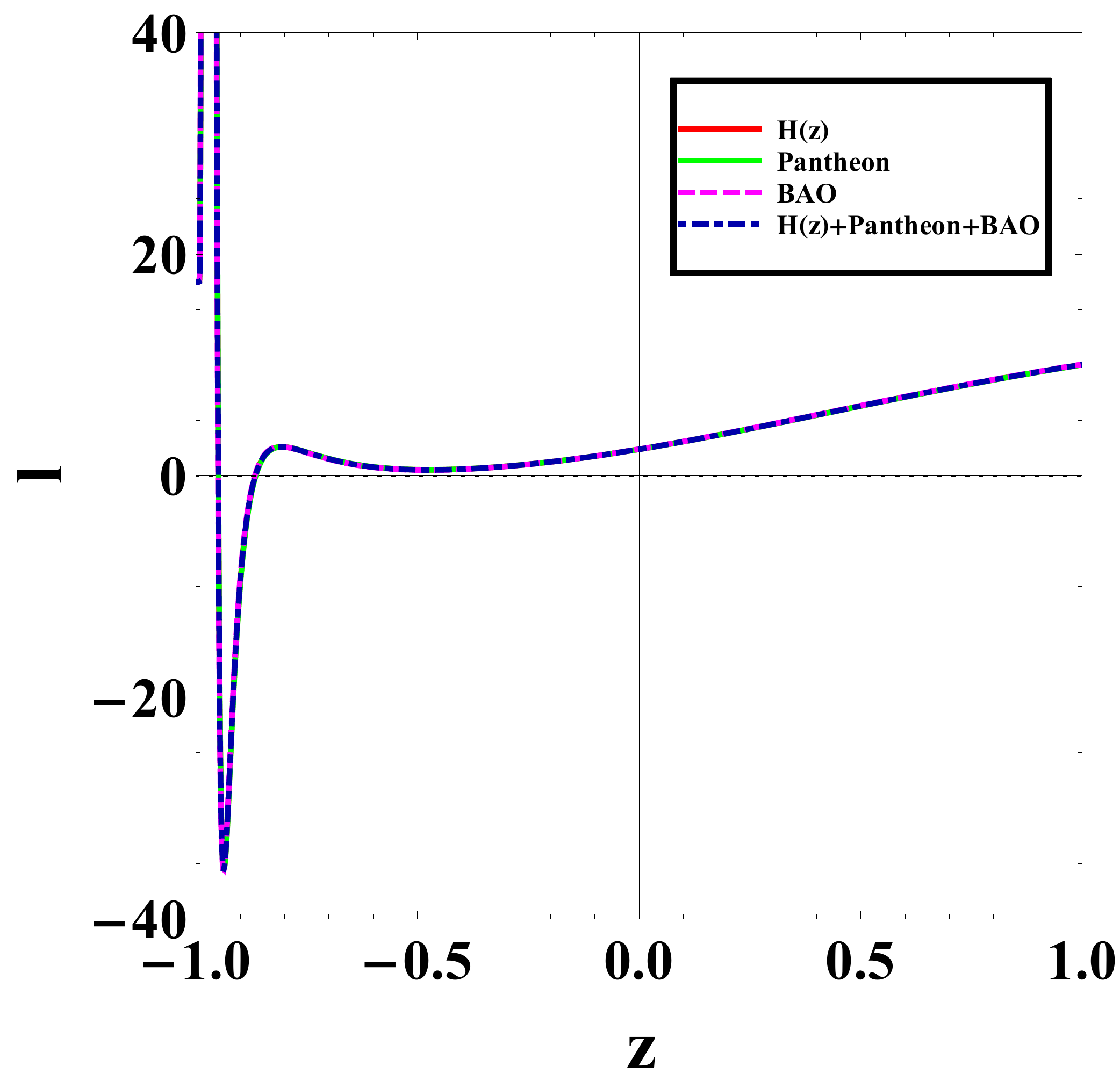}} 
	 { \includegraphics[scale=0.28]{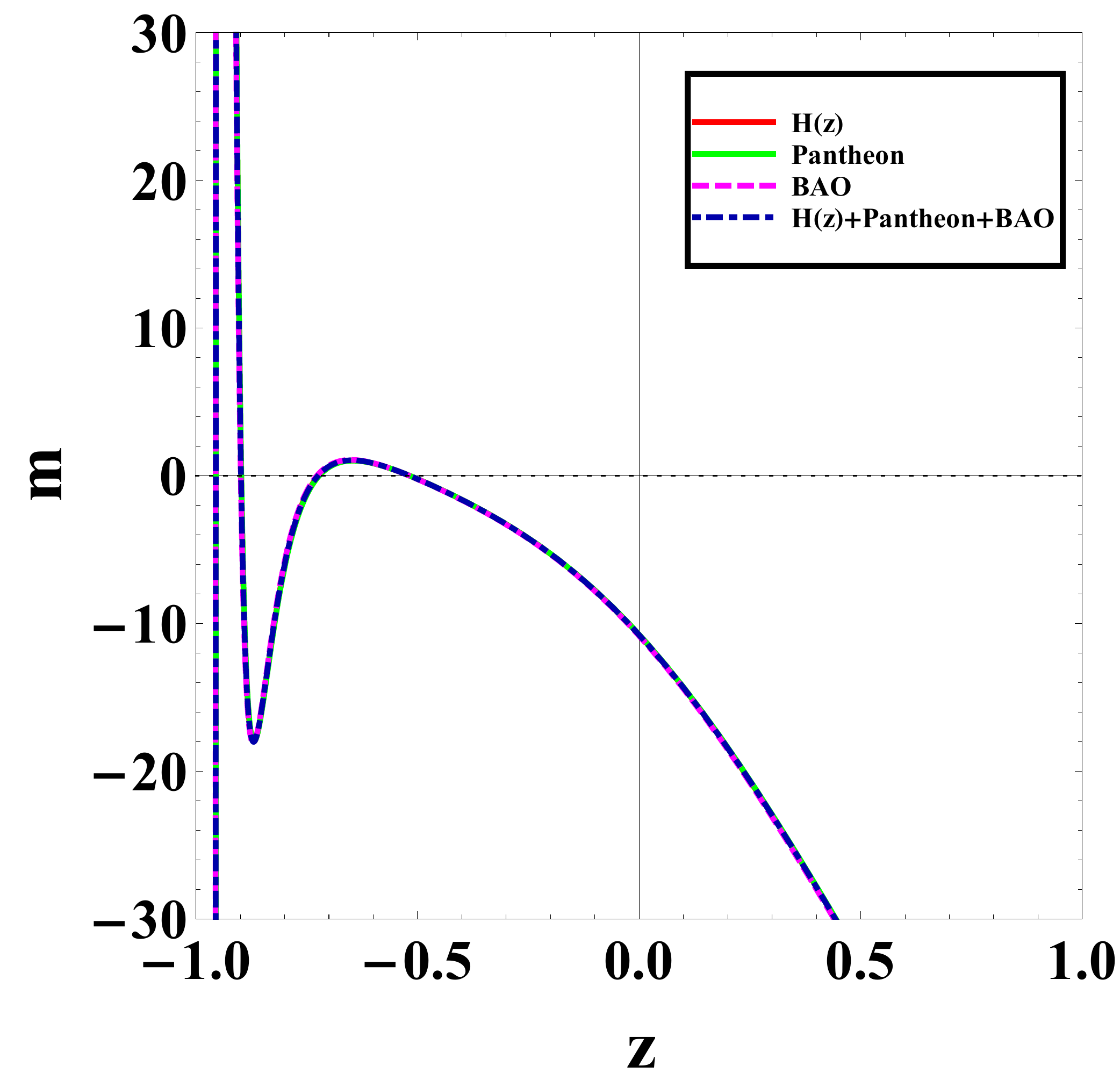}}
	\caption{{\it{  The evolution of  the   cosmographic parameters $j,s,l,m$ 
given in (\ref{39})-(\ref{39f}), in 
terms of the redshift $z$, using the 
best-fit values of the model parameters given in Tables 
\ref{tabparm1}-\ref{tabparm2}.  }} }
\label{s}
\end{figure}

We proceed to the examination of the other cosmographic parameters given in Eqs.
(\ref{39})-(\ref{39f}). In particular, we use the 
best-fit values of the model parameters given in Tables 
\ref{tabparm1}-\ref{tabparm2}, and in Fig. \ref{s} we present their evolution. 
Additionally, in Table \ref{tabparm3} we summarize their values at present.  
Since in the $ \Lambda$CDM paradigm, the value of the jerk 
parameter is equal to unity ($ j=1 $), the deviation from $ j=1 $ quantifies 
the deviation of a dark-energy scenario from the concordance model. Again we 
find that the new proposed dark-energy parametrization behaves similarly to $ 
\Lambda$CDM scenario at high redshifts, while the deviation becomes more 
significant at late and recent times, and especially in the future. Finally, 
the same features can be obtained from the evolution of the snap $ s $, lerk 
$ l $, and $ m $ parameters.

\begin{table}[ht]
\caption{ Summary of the constraints on the present values of the  cosmographic 
parameters, namely jerk $j$, snap $ s $, lerk 
$ l $ and $ m $ parameters, as well as on the present values of the  
statefinder diagnostic parameters $r$ and $s^*$. } \label{tabparm3}
\begin{center}
\begin{tabular}{l c c c c c c r} 
\hline\hline
{Dataset}  &  \,\,\,\,\,\, $ j $ \,\,\,\, \,\,\,\,\,\, & 
\,\,\,\,\,\, $ s $\,\,\,\, \,\,\,\,\,\ & \,\,\,\,\,\,\,\,\, $ l $\,\,\,\, 
\,\,\,\,\,\, &  \,\,\,\,\,\, $ m $ & \,\,\,\,\,\,\,    $ 
r $ \,\,\,\,\,\,\,\,\,\,   & \,\,\,\,\,\,\,\,\, $ s^* $ 
\,\,\,\,\,\,\,\,\,\, 
\\
\hline 
{ \scriptsize{ $ H(z)$}}
{ \scriptsize{(77 points data)}}    &   $ 0.506092 $ & $ 
-1.02864 $   &  $ 2.37997 $   &  $ -10.7706 $    & $ 
0.506^{+0.341}_{-0.269}  $     &   $0.174 ^{+0.124}_{-0.123} $
\\
{\scriptsize{$ \ \  Pantheon $ } }     &  $ 0.504616 $  &  $ -1.02915 $  &  $2.37783 $  &  $ -10.7529  $ &  $ 
0.505^{+0.238}_{-0.200} $ &   $ 0.174 ^{+0.074}_{-0.076}$
\\
{ \scriptsize{ $ BAO $}}
  &  $ 0.51083$  &  $ -1.02745 $  &  $2.38876 $   &  $ -10.8356 $  &  $ 0.511 
^{+0.227}_{-0.201}$     &   $ 0.172^{+0.074}_{-0.073} $
\\
{ \scriptsize{$ H(z) $ + $ Pantheon$}}
{ \scriptsize{ + $ BAO $ }}      &  $ 0.507417 $  &  $-1.02752 $  &  $ 2.37906 $   &  $ -10.7734 $  &  $ 
0.508^{+0.225}_{-0.198} $    &  $ 0.1729^{+0.073}_{-0.074} $
\\
\hline\hline  
\end{tabular}    
\end{center}
\end{table}

Let us now come to the statefinder diagnostic, which is based on   higher 
derivatives of the scale factor \cite{Sami:2012uh,Singh:2015hva, Myrzakulov:2013owa,Rani:2014sia}. In particular, one 
introduces a pair of geometrical parameters $ \{r,s^*\} $ in order to examine  
the dynamics of different     dark-energy models \cite{sah, ala}. The pair 
of parameters $ \{r,s^*\} $ are defined as: 
\begin{equation}\label{56}
r=\frac{\dddot{a}}{aH^{3}}\text{, \ \ }s^*=\frac{r-1}{3(q-\frac{1}{2})},
\end{equation}%
with $ q\neq \frac{1}{2}$.
For our  parametrization    (\ref{de-evol}), the 
expression for $ r $ is found to be
\begin{equation}\label{56a}
r= \frac{r_1+r_2+r_3}{(1 + z)^2 [2 + z (2 + z)]^2 
\tan^{-1}(1 + z)^2 \left[\pi (1 + z)^3 
\Omega_{b_0} + 4 e^{\frac{\eta z}{1+z}} \Omega_{x_0} \tan^{-1}(1 + 
z)^\eta\right]},
\end{equation}
where
\begin{eqnarray}
&&r_1=\pi (1 + z)^5 [2 + z (2 + z)]^2 \Omega_{m_0} \tan^{-1}(1 + z)^2 + 2 
e^{\frac{\eta z}{1+z}}(1 + z)^4 ( \eta-1) \eta\Omega_{x_0} \tan^{-1}(1 + 
z)^\eta, \nonumber\\
&&r_2=4 e^{\frac{\eta z}{1+z}}\eta\left\{ 2 \eta -3 + z [ 2 \eta -7 + z (\eta - 
2 z  -6 
)]\right\}\Omega_{x_0} \tan^{-1}(1 + z)^{1+\eta},
\nonumber\\
&&r_3=2 e^{\frac{\eta z}{1+z}}[2 + z (2 + z)]^2 [2 (1 + z)^2 - 4 (1 + z) \eta + 
\eta^2] \Omega_{x_0}\tan^{-1}(1 + z)^{2+\eta}.
\end{eqnarray}
Finally,  the expression for $s^*$ is obtained using (\ref{23}) and (\ref{56a}).

\begin{figure}[ht]
\centering
 {\includegraphics[scale=0.34]{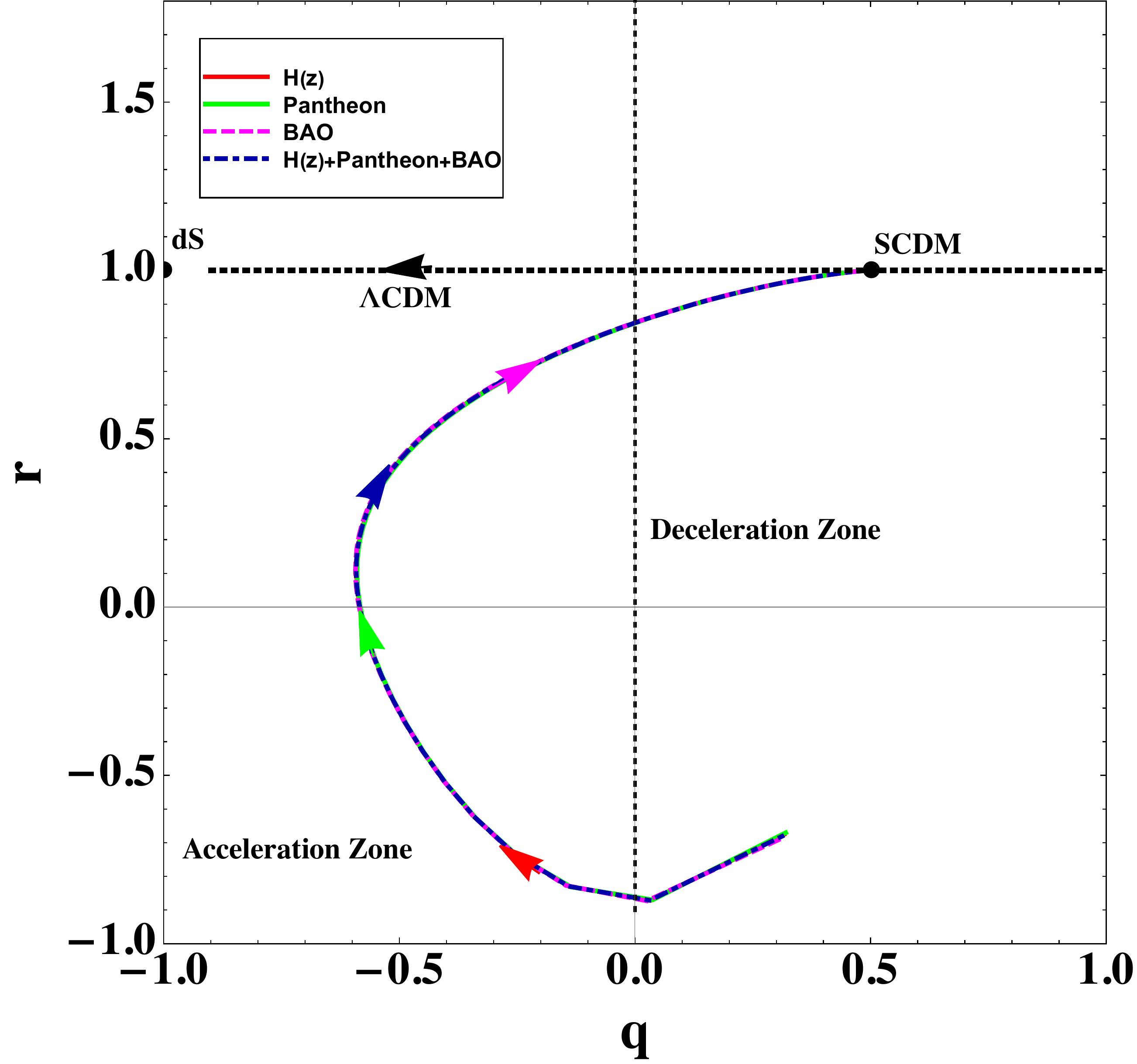}} 
 \caption{ {\it{Statefinder diagnostic trajectories in the $ q-r $ 
plane, using the 
best-fit values of the model parameters given in Tables 
\ref{tabparm1}-\ref{tabparm2}. }}}
\label{sr}
\end{figure}

In Fig. \ref{sr} we present trajectories for different 
observational datasets in the $ q-r $ plane. As we can see, all trajectories 
start from the decelerating zone, enter into the accelerating zone behaving 
close to $ \Lambda$CDM at present, and in the far future they converge 
to the CDM model without cosmological constant (namely $SCDM$ model) without 
resulting to the de Sitter ($ dS $) phase. The present values of parameters $ 
\{r,s^*\} $ of statefinder diagnostic are also given in Table \ref{tabparm3}. As we can see, the new parametrization at hand at high redshifts behaves as $\Lambda$CDM, while the deviation appears at small-redshifts and present time.

\section{Conclusions}
\label{Conclusions}

In this work, we proposed a novel dark-energy equation-of-state parametrization, with a single parameter $\eta$ that quantifies the deviation 
from $\Lambda$CDM cosmology. Firstly, we confronted the scenario with various datasets, from  Hubble function (OHD), Pantheon dataset, baryon 
acoustic oscillations (BAO) observations, and their joint dataset, and we presented the corresponding likelihood contours. As we saw, the new model parameter $\eta$ has a preference for a non-zero value, namely, a deviation from $\Lambda$CDM cosmology is favored, although the zero 
value is marginally inside the 1$\sigma$ confidence level. However, 
interestingly enough, we find that $H_0$  acquires a higher value compared to 
$\Lambda$CDM scenario, which implies that the new dark-energy parametrization 
at hand can partially alleviate the $H_0$ tension. 
 
Additionally, we performed a cosmographic analysis, examining the cosmographic 
parameters, namely deceleration $q$, jerk $j$, snap $ s $, lerk 
$ l $, and $ m $ parameters. As we showed, in the scenario at hand the universe transits from deceleration to acceleration in the recent 
cosmological past, however in the future it will not result in a de Sitter 
phase, since a second transition (at  $z_{tr_2} \approx -0.9$) will lead from 
acceleration to deceleration. Additionally, we found that the scenario  behaves
similarly to the $ \Lambda$CDM paradigm at high redshifts, while the deviation 
becomes significant at late and recent times (and thus the $H_0$ tension is 
alleviated), and especially in the future.  

Finally, we performed a statefinder diagnostic analysis. As we saw, all trajectories start from the decelerating 
zone, enter into the accelerating zone behaving close to $ \Lambda$CDM at 
present time,  while  in the far future, they converge to deceleration without 
resulting in the de Sitter phase. 

In summary, the new parametrization of the dark-energy equation of state 
with a single parameter is efficient in describing the data, as well as it can 
alleviate the $H_0$ tension. Hence, it would be worthy to proceed to more 
detailed investigations, such as to examine it at the perturbation level, and in 
particular relating to the $\sigma_8$ tension. Such an analysis will be 
performed in a separate project. 
 
\subsection*{Acknowledgements}
 J. K. Singh wishes to thank  M. Sami and S. G. Ghosh, for fruitful discussions. The authors also express their thanks to the referees for their valuable comments and suggestions.

\section{Observational data}
\label{AppendixA}
  
  In this Appendix, we present the observational datasets we use in our 
analysis, and we provide the relevant methodology and the 
 corresponding $\chi^2$.

\subsection{$H(z)$ data} 

In the case of observational Hubble data (OHD)  
   the corresponding $\chi^2$ of the  
maximum likelihood analysis is given by
\begin{equation}\label{29}
\chi _{OHD}^{2}(\eta, H_0)=\sum\limits_{i=1}^{77}\left[\frac{H_{th}(\Omega_{m0},\Omega_{x 0},H_0,\eta,z_{i})-H_{obs}(z_{i})}{\sigma _{H(z_i)}}\right] ^2,
\end{equation}
where $ H(z_i) $ is evaluated at redshift $ z_{i} $, while   $ H_{th} $ and $ 
H_{obs}$ represent the theoretical and observed value, and  $ \sigma_{H(z_{i})} 
$ is the standard deviation.  The detailed 
$H(z)$ data, namely the 77 points, are given in     Table \ref{tabff} below.

\begin{table}
\caption{The 77 Hubble Parameter Data from $ H(z)$ measurements used in the 
present analysis  in units of $\mathrm{km\,s^{-1}Mpc^{-1}}$. Method
 (a) corresponds to the Cosmic chronometric method, method  (b) to $ BAO $ signal 
in 
galaxy distribution, and method (c) to $ BAO $ signal in $ Ly\alpha $ forest 
distribution alone, or cross-correlated with $ QSOs $.} \label{tabff}
\begin{center}
\label{hubble}
\begin{tabular}{cccc}
\hline\hline
~~$z$ & ~~~~$H(z)$ $ (km/s/Mpc) $&~~~~ Method & ~~ Reference\\
\hline
0.00&~~	    69.1$ \pm $ 1.3&~~~~~~	  a&~~	    \cite{Farooq}\\  
0.07&~~  	70.4$ \pm $ 20&~~~~~~     a&~~	    \cite{Zhang} \\  
0.07&~~     69.0$ \pm $ 19.6&~~~~~~   a&~~      \cite{Zhang} \\  
0.09&~~  	70.4$ \pm $ 12.2&~~~~~~   a&~~      \cite{Simon}\\  
0.10&~~  	70.4$ \pm $ 12.2&~~~~~~   a&~~      \cite{Zhang}\\  
0.120&~~    68.6$ \pm $ 26.2&~~~~~~   a&~~      \cite{Zhang}\\   
0.12&~~  	70.0$ \pm $ 26.7&~~~~~~   a&~~      \cite{Farooq}\\   
0.170&~~    83.0$ \pm $ 8&~~~~~~      a&~~      \cite{Simon}\\  
0.170&~~ 	84.7$ \pm $ 8.2&~~~~~~    a&~~	    \cite{Simon}\\  
0.179&~~	76.5$ \pm $ 4&~~~~~~      a&~~      \cite{Moresco}\\  
0.1791&~~   75.0$ \pm $ 4&~~~~~~      a&~~      \cite{Moresco}\\  
0.199&~~	76.5$ \pm $ 5.1&~~~~~~	  a&~~      \cite{Moresco}\\  
0.1993&~~   75.0$ \pm $ 5&~~~~~~      a&~~       \cite{Moresco}\\    
0.200&~~    72.9$ \pm $ 29.6&~~~~~~   a&~~      \cite{Zhang}\\  
0.20&~~ 	74.4$ \pm $ 30.2&~~~~~~   a&~~      \cite{Zhang}\\  
0.24&~~ 	81.5$ \pm $ 2.7&~~~~~~	  b&~~      \cite{Gaztanaga}\\
0.27&~~ 	78.6$ \pm $ 14.3&~~~~~~   a&~~      \cite{Simon}\\  
0.280&~~    88.8$ \pm $ 36.3&~~~~~~   a&~~      \cite{Zhang}\\  
0.28&~~ 	90.6$ \pm $ 37.3&~~~~~~   a&~~      \cite{Farooq}\\   
0.35&~~ 	84.4$ \pm $ 8.6&~~~~~~    b&~~      \cite{Farooq}\\   
0.3519&~~   83.0$ \pm $ 14&~~~~~~     a&~~       \cite{Moresco}\\    
0.352&~~	84.7$ \pm $ 14.3&~~~~~~   a&~~       \cite{Moresco}\\   
0.38&~~ 	81.5$ \pm $ 1.9&~~~~~~	  b&~~      \cite{Alam}\\  
0.3802&~~   83.0$ \pm $ 13.5&~~~~~~   a&~~      \cite{Moresco1}\\ 
0.3802&~~	84.7$ \pm $ 14.1&~~~~~~   a&~~      \cite{Moresco1}  \\
0.40&~~     95.0$ \pm $ 17&~~~~~~     a&~~      \cite{Simon}\\    
0.40&~~  	96.9$ \pm $ 17.3&~~~~~~   a&~~      \cite{Simon}\\  
0.4004&~~   77.0$ \pm $ 10.2&~~~~~~   a&~~      \cite{Moresco1}\\
0.4004&~~	78.6$ \pm $ 10.4&~~~~~~   a&~~     \cite{Moresco1}\\
0.4247&~~   87.1$ \pm $ 11.2&~~~~~~   a&~~      \cite{Moresco1}\\ 
0.4247&~~	88.9$ \pm $ 11.4&~~~~~~	  a&~~      \cite{Moresco1} \\ 
0.43&~~ 	88.3$ \pm $ 3.8&~~~~~~	  a&~~      \cite{Farooq}\\   
0.44&~~ 	84.3$ \pm $ 7.9&~~~~~~	  a&~~     \cite{Blake}\\ 
0.4497&~~   92.8$ \pm $ 12.9&~~~~~~   a&~~      \cite{Moresco1}\\ 
0.4497&~~	94.7$ \pm $ 13.1&~~~~~~   a&~~      \cite{Moresco1}\\  
0.470&~~    89.0$ \pm $ 34.0&~~~~~~   a&~~      \cite{Ratsimbazafy}\\   
0.47&~~ 	90.8$ \pm $ 50.6&~~~~~~   a&~~      \cite{Ratsimbazafy}\\ 
0.4783&~~   80.0$ \pm $ 99.0&~~~~~~   a&~~      \cite{Moresco1}\\  
0.4783&~~	82.5$ \pm $ 9.2&~~~~~~	  a&~~      \cite{Moresco1}\\  	
0.48&~~ 	99.0$ \pm $ 63.2&~~~~~~   a&~~      \cite{Ratsimbazafy}\\   
0.51&~~ 	90.8$ \pm $ 1.9&~~~~~~	  b&~~      \cite{Alam}\\  
0.57&~~ 	98.8$ \pm $ 3.4&~~~~~~	  b&~~      \cite{Zhang}\\  
0.593&~~    104.0$ \pm $ 13.0&~~~~~~  a&~~       \cite{Moresco}\\    
0.593&~~	106.1$ \pm $ 13.3&~~~~~~   a&~~       \cite{Moresco}\\   
0.60&~~     89.7$ \pm $ 6.2&~~~~~~    a&~~      \cite{Zhang}\\  
0.61&~~ 	97.8$ \pm $ 2.1&~~~~~~    b&~~      \cite{Alam}\\ 
0.64&~~     98.82$ \pm $ 2.98&~~~~~~   b&~~     \cite{Wang1}\\
\hline\hline
\end{tabular}
\label{table1}
\end{center}
\end{table}

\begin{table}
\begin{center}
\label{hubble}
\begin{tabular}{cccc}
\hline\hline
~~$z$ & ~~~~$H(z)$ $ (km/s/Mpc) $&~~~~ Method & ~~ Reference\\
\hline
0.6797&~~   92.0$ \pm $ 8&~~~~~~      a&~~       \cite{Moresco}\\  
0.68&~~ 	93.9$ \pm $ 8.1&~~~~~~    a&~~	     \cite{Moresco}\\ 
0.73&~~     99.3$ \pm $ 7.1&~~~~~~    a&~~     \cite{Blake}\\  
0.7812&~~   105.0$ \pm $ 12&~~~~~~    a&~~       \cite{Moresco}\\      
0.781&~~	107.1$ \pm $ 12.2&~~~~~   a&~~       \cite{Moresco}\\   
0.875&~~	127.6$ \pm $ 17.3&~~~~~   a&~~       \cite{Moresco}\\  
0.8754&~~   125.0$ \pm $ 17&~~~~~~    a&~~       \cite{Moresco}\\  
0.88&~~ 	91.8$ \pm $40.8&~~~~~~    a&~~       \cite{Stern}\\  
0.880&~~    90.0$ \pm $ 40&~~~~~~     a&~~      \cite{Ratsimbazafy}\\  
0.90&~~     69.0 $ \pm $ 12&~~~~~~    a&~~      \cite{Simon}\\     
0.90&~~  	119.4$ \pm $ 23.4&~~~~~   a&~~      \cite{Simon}\\
0.900&~~    117.0$ \pm $ 23&~~~~~~    a&~~      \cite{Simon}\\  
1.037&~~	157.2$ \pm $ 20.4&~~~~~~  a&~~       \cite{Moresco}\\  
1.037&~~    154.0$ \pm $ 20&~~~~~~    a&~~       \cite{Moresco}\\ 
1.30&~~	    171.4$ \pm $ 17.3&~~~~~	  a&~~      \cite{Simon}\\  
1.300&~~    168.0$ \pm $ 17&~~~~~~    a&~~      \cite{Simon}\\ 
1.363&~~    160.0$ \pm $ 33.6&~~~~~~  a&~~      \cite{Moresco2}\\  
1.363&~~	163.3$ \pm $ 34.3&~~~~~	  a&~~      \cite{Moresco2}\\  
1.430&~~    177.0$ \pm $ 18&~~~~~~    a&~~      \cite{Simon}\\   
1.43&~~	    180.6$ \pm $ 18.3&~~~~~~  a&~~      \cite{Simon}\\  
1.530&~~    140.0$ \pm $ 14&~~~~~~    a&~~      \cite{Simon}\\  
1.53&~~	    142.9$ \pm $ 14.2&~~~~~~  a&~~      \cite{Simon}\\  
1.750&~~    202.0$ \pm $ 40&~~~~~~    a&~~      \cite{Simon}\\  
1.75&~~	    206.1$ \pm $ 40.8&~~~~~~  a&~~      \cite{Simon}\\  
1.965&~~    186.5$ \pm $ 50.4&~~~~~~  a&~~      \cite{Moresco2}\\  
1.965&~~	190.3$ \pm $ 51.4&~~~~~~  a&~~      \cite{Moresco2}\\ 
2.30&~~	    228.0$ \pm $ 8.1&~~~~~~	  c&~~	    \cite{Delubac}\\  
2.34&~~	    226.5$ \pm $ 7.1&~~~~~~	  c&~~	    \cite{Delubac}\\ 
2.36&~~	    230.6$ \pm $ 8.2&~~~~~~	  c&~~      \cite{Font-Ribera}\\  
\hline\hline
\end{tabular}
\end{center}
\label{table2}
\end{table}
 
\subsection{ Pantheon Data}
 
 We use the latest published dataset for supernovae type $Ia$ the Pantheon sample, consisting of 1048 data points. In our analysis, we utilize these data points, which have been confirmed spectroscopically by SNe$Ia$ and cover the redshift range of $0.01 < z < 2.26$.  The $\chi^2_{Pantheon}$ function for the Pantheon dataset is taken as.  

\begin{equation}\label{30}
    \chi_{Pan}^{2}=\sum\limits_{i=1}^{1048}\left[ \frac{\mu_{th}(\Omega_{m0},\Omega_{x 0},H_0,\eta,z_{i})-\mu_{obs}(z_{i})}{\sigma _{\mu(z_{i})}}\right] ^2,
\end{equation}
where $ \mu_{th} $ and $ 
\mu_{obs} $ are the theoretical and observed distance modulus, and  
$\sigma_{\mu(z_{i})}$
the standard deviation. 
The distance modulus  $\mu(z)$ is defined as 
\begin{equation}\label{31}
\mu(z)= m-M = 5Log D_l(z)+\mu_{0},
\end{equation}
where $ m $ and $ M $ denote the apparent and absolute magnitudes. 
Additionally, the luminosity distance $ D_l(z) $ for flat Universe and the 
nuisance parameter $ \mu_0 $ are given by
\begin{equation}\label{32}
D_l(z)=(1+z)H_0\int_0^z \frac{1}{H(z^*)}dz^*,
\end{equation}
and
\begin{equation}\label{33}
\mu_0= 5Log\Big(\frac{H_0^{-1}}{1Mpc}\Big)+25,
\end{equation}
respectively.  

\subsection{ Baryon acoustic oscillations (BAO)} 

Concerning  Baryon Acoustic Oscillations (BAO) we use the data from   Sloan 
Digital Sky Survey ($SDSS $) \cite{padn}, $ 6dF $ Galaxy survey ($ 6dFGS $) 
\cite{bue}, $ BOSS\, CMASS $ \cite{boss} and three parallel measurements from $ 
WiggleZ $  survey 
\cite{wig}.
In  BAO  observations the distance redshift ratio $ d_z $ is 
\begin{equation}\label{34}
d_z=\frac{r_s(z^*)}{D_v(z)},
\end{equation}
where $ z^*=1090 $   is the redshift at the time of photon decoupling   
\cite{adep}, and  $ r_s(z^*) $ is the corresponding comoving sound horizon 
  \cite{waga}. The dilation scale defined by Eisenstein et al. 
\cite{eis} is  
\begin{equation}\label{34a}
D_v(z)= \left[(1+z)^2\frac{d^2_A(z) z}{H(z)}\right]^{\frac{1}{3}},
\end{equation}
where $ d_A(z) $ is the angular diameter distance, which  essentially is a 
geometric mean of two transverse and
one radial direction. The  value of $ \chi _{BAO}^{2} $ is given by \cite{gio}
\begin{equation}\label{35}
\chi _{BAO}^{2}= A^{T}C^{-1}A,
\end{equation}
where  
 \[
   A=
  \left[ {\begin{array}{cc}
   \frac{d_A(z^*)}{D_v(0.106)} - 30.95\\
       \frac{d_A(z^*)}{D_v(0.2)} - 17.55\\
       \frac{d_A(z^*)}{D_v(0.35)} - 10.11\\
       \frac{d_A(z^*)}{D_v(0.44)} - 8.44\\
       \frac{d_A(z^*)}{D_v(0.6)} - 6.69\\ 
       \frac{d_A(z^*)}{D_v(0.73)} - 5.45\\
  \end{array} } \right],
\]
and the inverse covariance matrix $ C^{-1} $ is
\[
  C^{-1}=
  \left[ {\begin{array}{cccccc}
   0.48435 & -0.101383 & -0.164945 & -0.0305703 & -0.097874 & -0.106738\\
 -0.101383 & 3.2882 & -2.45497 & -0.0787898 & -0.252254 & -0.2751 \\
 -0.164945 & -2.454987 & 9.55916 & -0.128187 & -0.410404 & -0.447574\\
-0.0305703 & -0.0787898 & -0.128187 & 2.78728 & -2.75632 & 1.16437 \\
-0.097874 & -0.252254 & -0.410404 & -2.75632 & 14.9245 & -7.32441 \\
-0.106738 & -0.2751 & -0.447574 & 1.16437 & -7.32441 & 14.5022 \\
  \end{array} } \right],
  \]
approaching the correlation coefficients available in \cite{hing, gio}. 

\subsection{Joint analysis}

In the case where some of the above datasets are used simultaneously, the 
corresponding $\chi^2$ arises from the sum of the separate ones. 
In particular, 
we will use the following combinations:

\begin{equation}\label{37}
\chi_{HPB}^{2}=\chi _{OHD}^{2}+\chi _{Pan}^{2}+\chi_{BAO}^{2},
\end{equation}

\end{document}